\documentclass{INTERSPEECH2023}

\interspeechcameraready

\usepackage{hyperref}
\usepackage[hyphenbreaks]{breakurl}

\usepackage{enumitem}
\setlist[itemize]{align=parleft,left=0pt..1em}

\usepackage{siunitx}
\usepackage{listings}
\usepackage{xcolor}

\usepackage{algorithm}
\usepackage{algpseudocode}

\usepackage{amsmath,graphicx}
\usepackage{color}
\usepackage{setspace}
\hypersetup{
   breaklinks=true,   
   pdfusetitle=true,  
}

\usepackage{color}
\definecolor{codegreen}{rgb}{0,0.6,0}
\definecolor{codegray}{rgb}{0.5,0.5,0.5}
\definecolor{codepurple}{rgb}{0.58,0,0.82}
\definecolor{backcolour}{rgb}{1.00,1.00,1.00}
\lstdefinestyle{mystyle}{
    backgroundcolor=\color{backcolour},   
    commentstyle=\color{codegreen},
    keywordstyle=\color{magenta},
    numberstyle=\tiny\color{codegray},
    stringstyle=\color{codepurple},
    basicstyle=\ttfamily\footnotesize,
    breakatwhitespace=false,         
    breaklines=true,                 
    captionpos=b,                    
    keepspaces=true,                 
    numbers=left,                    
    numbersep=5pt,                  
    showspaces=false,                
    showstringspaces=false,
    showtabs=false,                  
    tabsize=2
}
\title{2-bit Conformer quantization for automatic speech recognition}
\name{Oleg Rybakov, Phoenix Meadowlark, Shaojin Ding, David Qiu, 
Jian Li, David Rim, Yanzhang He}

\address{Google Research}
\email{\{rybakov,meadowlark,shaojinding,qdavid,jianlijianli,davidrim,yanzhanghe\}@google.com}

\usepackage{setspace}

\begin{document}

\setstretch{0.92}

\maketitle
 
\begin{abstract}
Large speech models are rapidly gaining traction in research community. As a result, model compression has become an important topic, so that these models can fit in memory and be served with reduced cost. Practical approaches for compressing automatic speech recognition (ASR) model use int8 or int4 weight quantization. In this study, we propose to develop 2-bit ASR models. We explore the impact of symmetric and asymmetric quantization combined with sub-channel quantization and clipping on both LibriSpeech dataset and large-scale training data. We obtain a lossless 2-bit Conformer model with 32\% model size reduction when compared to state of the art 4-bit Conformer model for LibriSpeech. With the large-scale training data, we obtain a 2-bit Conformer model with over 40\% model size reduction against the 4-bit version at the cost of 17\% relative word error rate degradation.
\end{abstract}

\section{Introduction}
\label{sec:intro}






Modern automatic speech recognition models are mostly based on an end-to-end approach~\cite{li2020comparison,CC18,KimHoriWatanabe17,JinyuLi2019}. One of the popular methods to improve accuracy of such models is to increase model size~\cite{gulati2020conformer}. With the growing success and size of these models,  compressing them with neutral quality impact is becoming an important research topic.

The most popular approaches for compressing neural networks are pruning~\cite{strom1997sparse, takeda2017node}, knowledge distillation~\cite{li2018compression}, and quantization~\cite{han2015deep, alvarez2016efficient}. In this paper we are focused on quantization as the most straight forward approach. It can be applied on activations and weights. If both activations and weights are quantized~\cite{quantize_weight_act}, then it reduces memory footprint and can give speed up due to memory footprint reduction and low bits multiplication (the latter one requires hardware support). If quantization is applied only on weights then it reduces memory footprint and can provide speed up due to lower memory usage. It also does not require special hardware support for low bits numbers multiplication. That is why in this work we are focused on weights only quantization.

Quantization methods can be divided into post training quantization (PTQ) and quantization aware training (QAT). PTQ is successfully applied on speech applications~\cite{he2019streaming, sainath2020streaming} because it is easy to use (e.g. with TFLite~\cite{POSTQUANT}) and it works well at int8 precision. PTQ with lower bits has limited support in TFLite~\cite{POSTQUANT} and can have significant accuracy degradation with no or minimal tools to address it. Hence, we are focusing on quantization aware training. QAT can be applied after model is pre-trained (i.e., fine-tuning stage), or it can be applied from the beginning of the model training (i.e., QAT training from scratch). In this work we are focused on training from scratch, although our approach can be used for fine-tuning too. Quantization of a tensor can be done with dynamic quantization~\cite{pact} or static quantization~\cite{clipping1, abdolrashidi2021pareto}. In this work we are focused on dynamic quantization because it does not require additional variables during training and it works well for speech applications e.g.~\cite{ASR4BITS}. Tensor can be quantized using non-uniform quantization (e.g., with float8~\cite{flt_quant}). Not all hardware supports it, so in this work we are focused on uniform 2-bit integer quantization, also called fixed-point quantization~\cite{clipping1}.

QAT with 4-bit is successfully applied on multiple speech models~\cite{hotword4bit, fasoli20214, ASR4BITS, parrot4bit, sub8bits} with minimal accuracy impact. Lower than 4-bit weight quantization is explored for different applications~\cite{low_bits_img0, low_bits_img, xnor_net}, but there is not much research on 2-bit quantization of ASR models. One work~\cite{asr1bit2bit4bit} addresses lower than 4-bit quantization, but the authors showed significant accuracy degradation with 2-bit and 4-bit quantization. Note that in \cite{asr1bit2bit4bit}, the authors quantize both activation and weights. Here we are focused on 2-bit Conformer weights only quantization with minimal accuracy impact. Our main contributions are outlined as below:

\begin{itemize}[leftmargin=*]
    \item We present a new 2-bit asymmetric dynamic sub-channel QAT technique with adaptive per channel clipping (based on greedy search). It is open sourced in Praxis~\cite{praxis}.
    \item We benchmark several proposed approaches of QAT and demonstrate that Conformer ASR on LibriSpeech data shows minimal or no accuracy loss with 2-bit weights when comparing to state of the art float model. We reduced model size by 32\% relative to the 4-bit model and establish state of the art ASR model in terms of model size and WER.
    \item We evaluate the effectiveness of the best 2-bit setup on a Conformer model that is trained on large-scale datasets. We achieve over 40\% model size reduction against the 4-bit version at the cost of 17\% relative word error rate degradation.
\end{itemize}

\section{Quantization aware methods}




\subsection{Symmetric quantization: \textit{I2Wsym}}
\label{sec:symmetric}

The standard method of weights only QAT is based on symmetric quantization~\cite{white_paper_quantization}. State of the art 4-bit symmetric quantization is presented in~\cite{ASR4BITS}. We use approach described in~\cite{ASR4BITS} as the baseline, configure it for 2-bit quantization, and label it as \textit{I2Wsym}. Note that 2-bit symmetric quantization under-utilizes the 2-bit quantization buckets (because it uses only three values). As a result, symmetric quantization can degrade accuracy.

\subsection{Asymmetric quantization: \textit{I2Wasym}}
\label{sec:asym}
\begin{figure}[t]
    \centering
    \includegraphics[width=0.9\columnwidth]{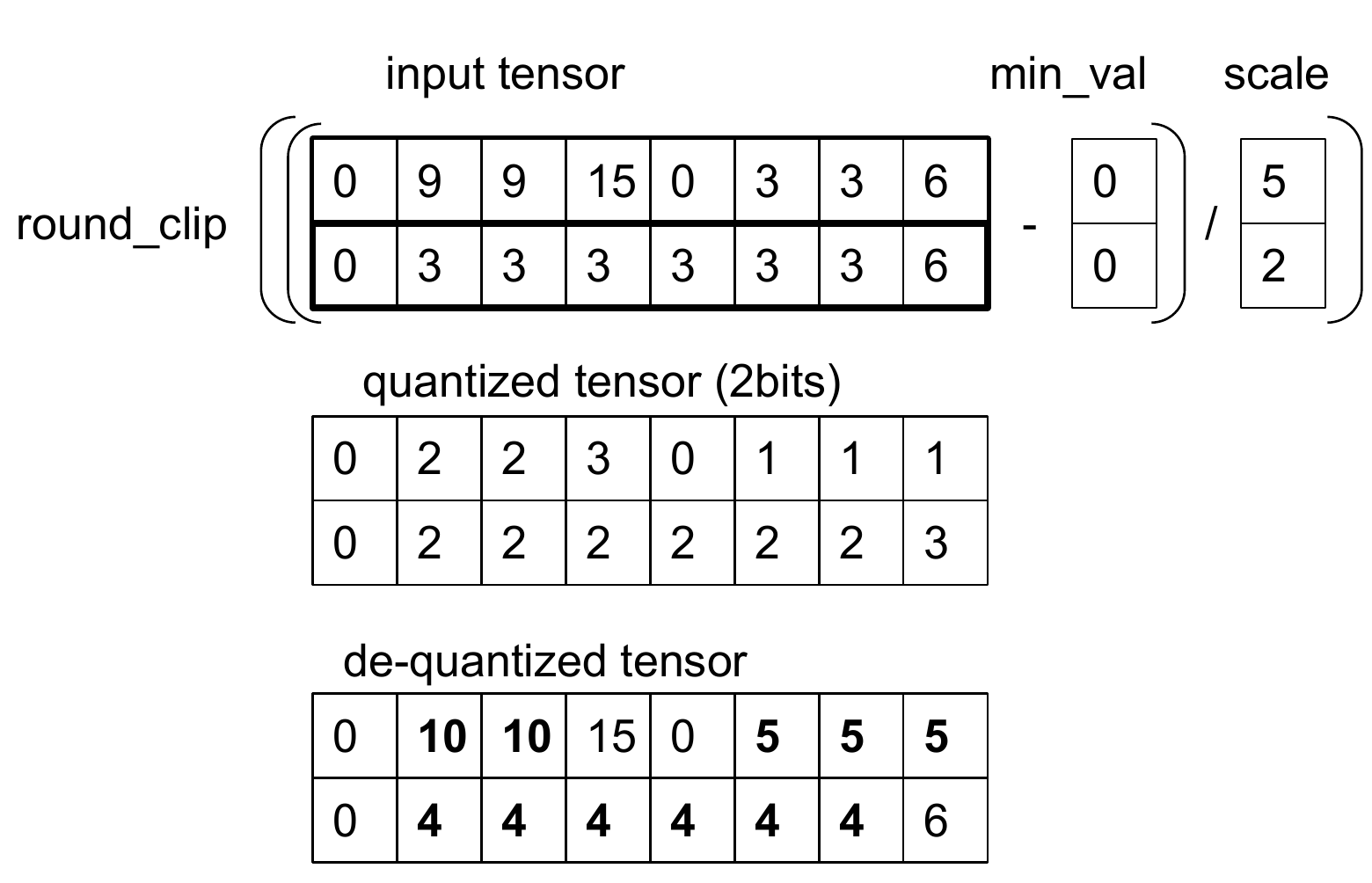}
    \caption{Per-channel asymmetric quantization.}
    \label{fig:per_channel}
    \vspace{-4mm}
\end{figure}

\begin{figure}[t]
    \centering
    \includegraphics[width=0.9\columnwidth]{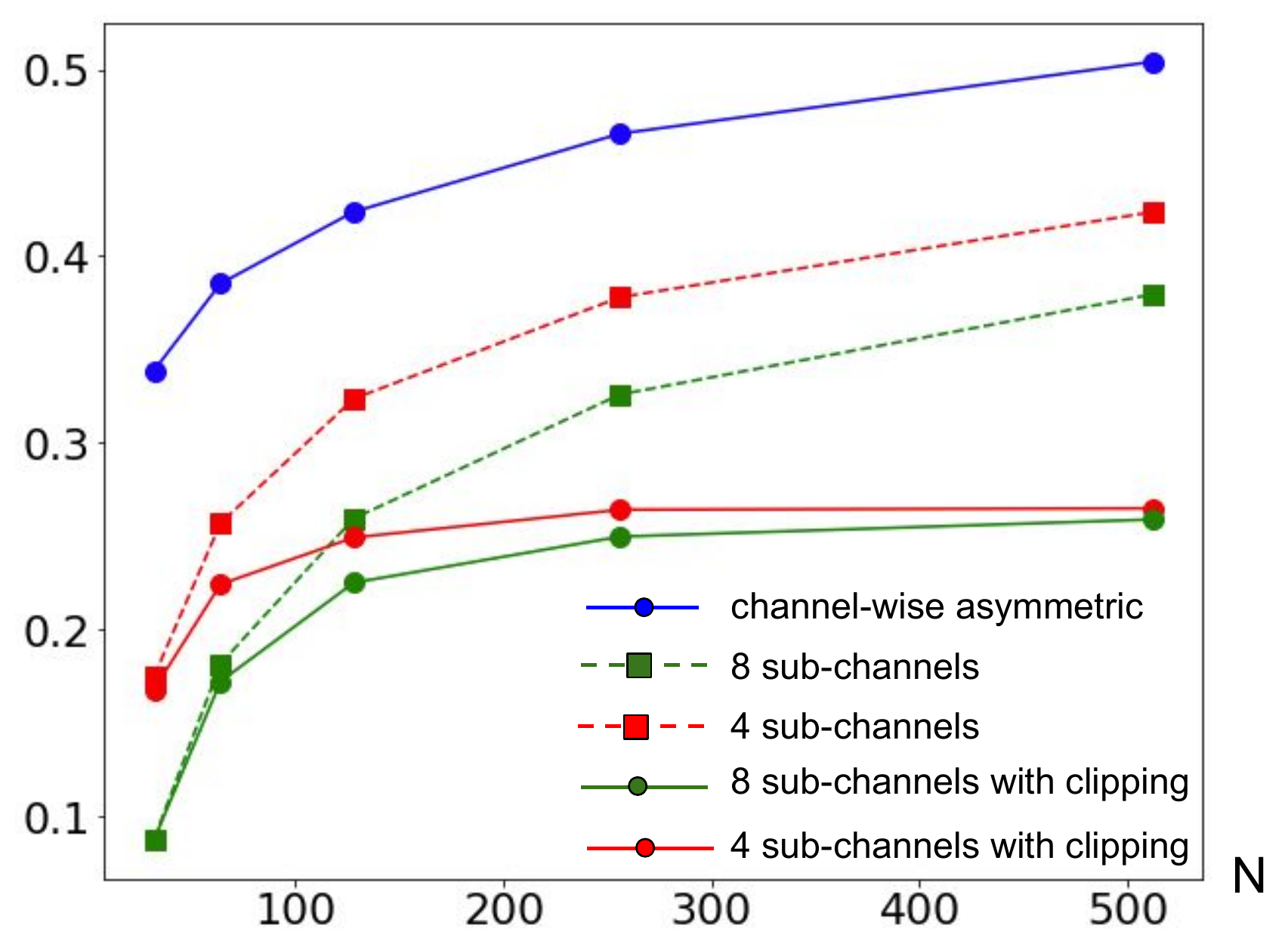}
    \caption{The per-entry mean absolute quantization error plotted against the size of the input dimension.}
    \label{fig:error_graph}
\end{figure}

Asymmetric quantization~\cite{white_paper_quantization} allows us to use all four quantization buckets by estimating the minimum value and subtracting it from the input tensor. We label 2-bit asymmetric quantization as \textit{I2Wasym}. In Figure\ref{fig:per_channel}, we show an example with per channel asymmetric quantization. The input tensor (weight matrix) has two rows (every row is called a channel). We quantize it using the function \textit{quantize} (shown on Figure~\ref{fig:quant_functions}). It computes min value (\textit{min\_val}) for every row (using function \textit{scale\_and\_min} from Figure~\ref{fig:quant_functions}) then subtracts it from input tensor and divides by scale. In Figure~\ref{fig:per_channel}, "quantized tensor" is the output of function \textit{round\_clip}, which rounds and clips the tensor between zero and three (according to 2-bit quantization range). The de-quantized tensor is shown at the bottom of Figure~\ref{fig:per_channel}. It is computed using the  \textit{dequantize} function (shown in Figure~\ref{fig:quant_functions}). As we can see in Figure~\ref{fig:per_channel}, most de-quantized values (highlighted by bold) are different from the original values in the input tensor. The reason for such quantization error is outliers in the input tensor. We use Figure~\ref{fig:per_channel} for illustration purposes. In Figure~\ref{fig:error_graph}, we show quantization error of asymmetrical per channel quantization (blue curve in Figure~\ref{fig:error_graph}) applied on an input tensor with size [32, N], where 32 is the number of channels and N can be 32, 64, 128, 256 or 512 (x-axis). The input tensor is filled with standard normal noise (with zero mean and unit variance), and then it is quantized and de-quantized. The quantization error (y-axis) is defined as the mean (over all entries) absolute difference between input tensor and corresponding de-quantized tensor.

\subsection{Asymmetric quantization with scale backpropagation: \textit{I2WasymSc}}
\label{sec:asym_sc}

In~\cite{ASR4BITS}, the authors used full Straight-Through Estimator(STE) for quantization  aware training of 4-bit ASR. We call it full STE because the gradient did not propagate through round and scale computation.
To improve model quality we enable gradient over scale~\cite{RAND} and set \textit{stop\_gradient} equal false in the function \textit{scale\_and\_min} in Figure~\ref{fig:quant_functions}. This approach allows to reduce outliers as described in~\cite{RAND}.
2-bit asymmetric quantization with scale backpropagation will be labeled as \textit{I2WasymSc}. All subsequent QAT methods will use backpropagation over scale as well.

\subsection{Asymmetric quantization with scale backpropagation, sub-channel and adaptive clipping: \textit{I2WasymScSubchClip}}
\label{sec:asym_sc_sub_clip}

\begin{figure}[t]
    \centering
    \includegraphics[width=0.7\columnwidth]{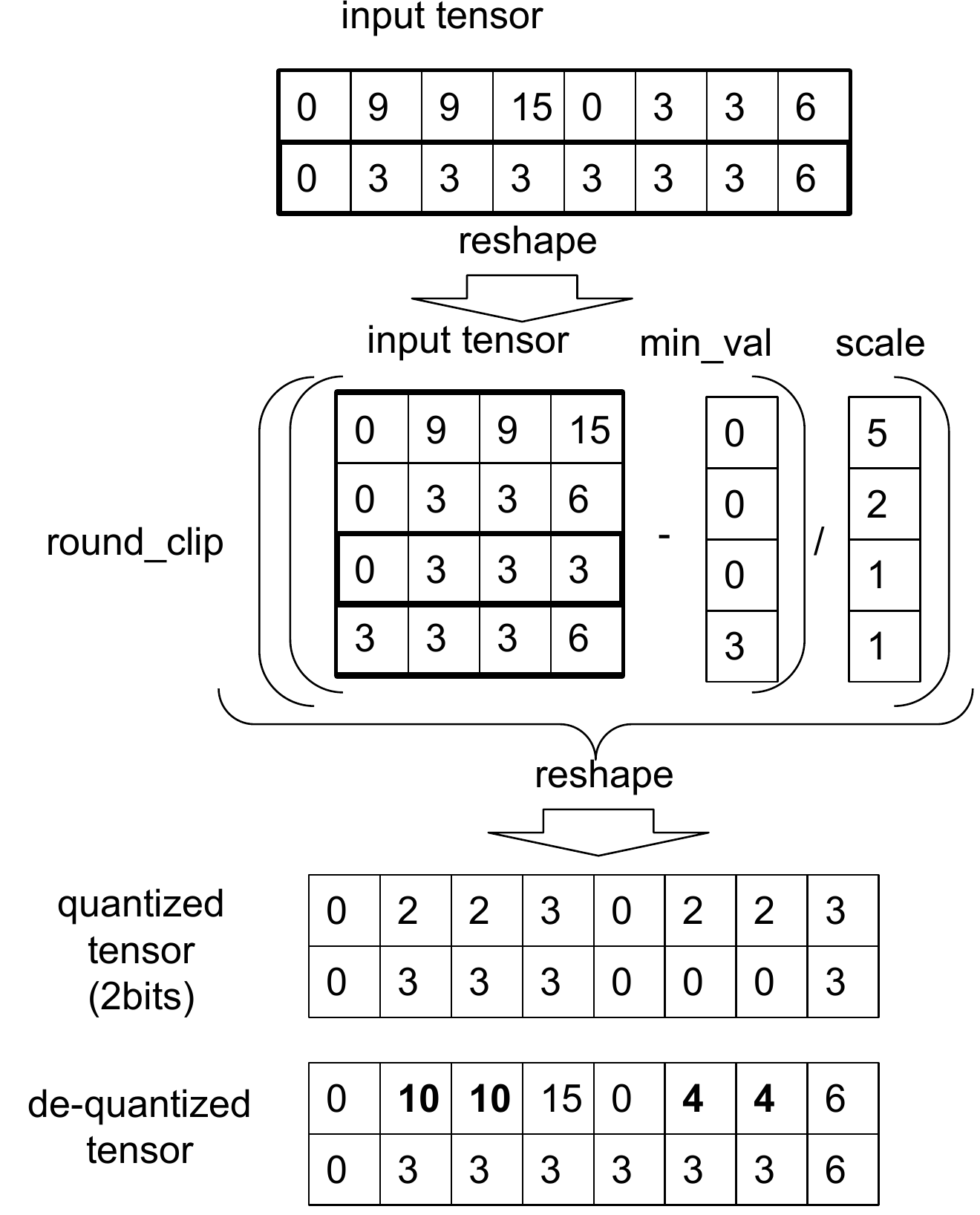}
    \caption{Sub-channel quantization.}
    \label{fig:sub_channel}
\end{figure}

One of the methods of dealing with the outliers in the input tensor (model weights) is based on sub-channel quantization~\cite{sub-channel}, which is also similar to group based quantization~\cite{group_quant, ptq_group}. The key idea is to split a channel into several sub-channels and then quantize them independently. It can introduce additional overhead because with more sub-channels, we will need to keep more quantization metadata: scales and minimum values. An example of such an approach is shown in Figure~\ref{fig:sub_channel}. The input tensor with two channels is reshaped, so that there are 4 sub-channels. Afterwards, the same quantization operations, described on Figure\ref{fig:per_channel}, are applied. As we can see, the de-quantized tensor has only four different values (highlighted by bold on Figure~\ref{fig:sub_channel}) when compared against the input tensor. In Figure~\ref{fig:error_graph} we show the quantization error of this approach when splitting into 4 sub-channels (red dashed line) and 8 sub-channels (green dashed line). As expected, the quantization error with 4 sub-channels is lower than that of channel-wise asymmetric quantization (described in section~\ref{sec:asym}). Furthermore, the quantization error becomes even lower after increasing the number of sub-channels to 8.

\begin{figure}[t]
    \centering
    \includegraphics[width=1.0\columnwidth]{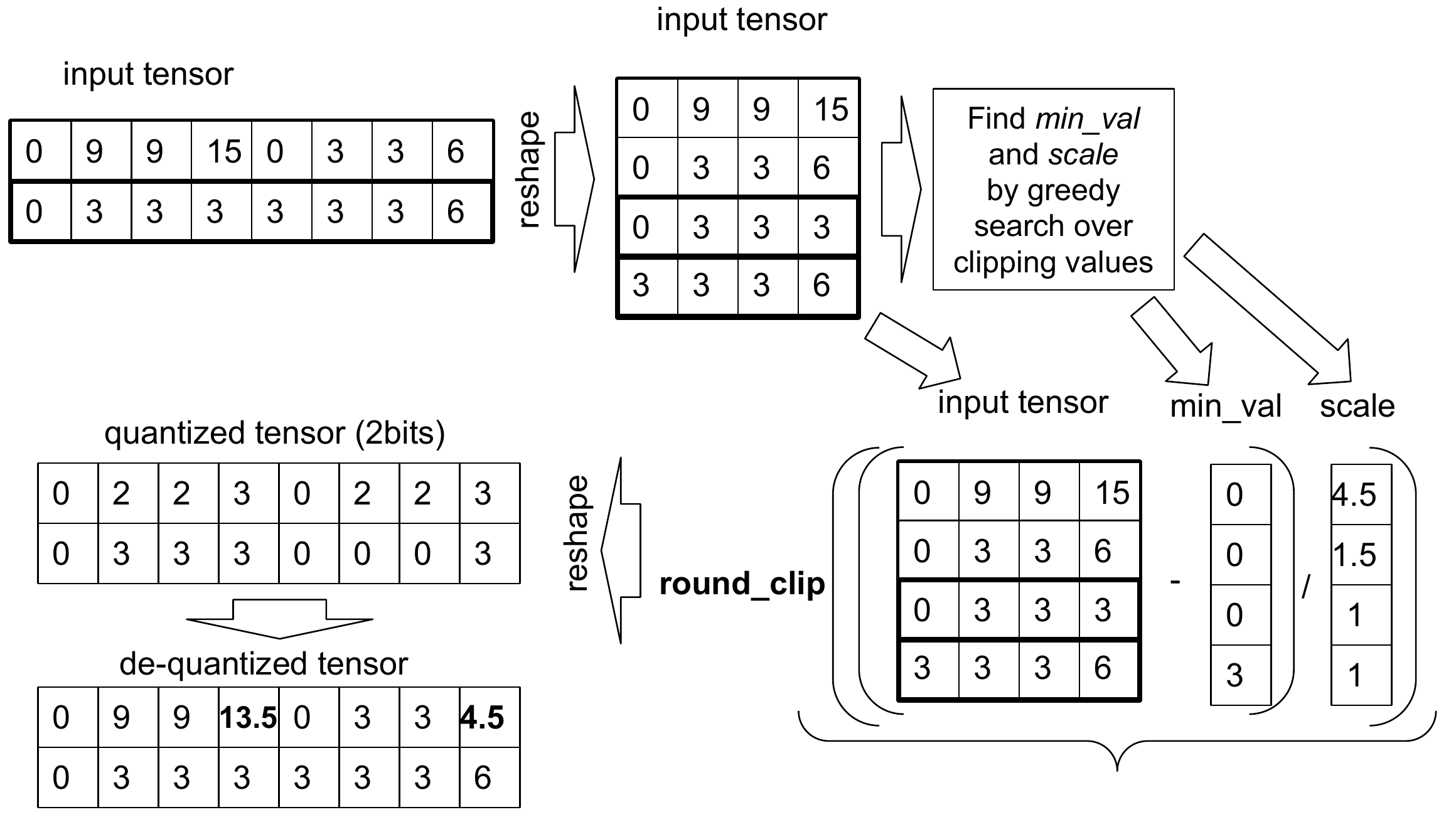}
    \caption{Sub-channel quantization with clipping per sub-channel.}
    \label{fig:sub_channel_per_tensor_clip}
    \vspace{-0.1in}
\end{figure}

Another approach to reduce the number of the outliers in the input tensor is based on clipping. For example, in PACT~\cite{pact}, the authors propose to learn clipping parameters to improve the quality of activation quantization. In~\cite{clipping1, clipping2}, the clipping value is estimated based on percentile. In~\cite{clipping2}, the authors design OCTAV~\cite{clipping2} algorithms for online estimation of clipping values. In this work, we propose to use greedy search for clipping parameter estimation. Hence, we combine sub-channel quantization with clipping as demonstrated on Algorithm~\ref{sub_channel_clipping}, and label it as \textit{I2WasymScSubchClip}. The input tensor (weight matrix) is reshaped so that the channel dimension is divided into several sub-channels. For example, in Figure~\ref{fig:sub_channel_per_tensor_clip}, two input channels are divided into four sub-channels. Then we run \textit{greedy\_search} over clipping values in the range of $[0.5, 1.0]$ with 0.05 step, and compute mean absolute error between the input weights and the quantized/de-quantized weights (for quantization and de-quantization we call functions presented on Figure~\ref{fig:quant_functions}). We apply different clipping values per sub-channel and select the quantized tensor with clipping values that correspond to the minimal quantization error. An example of quantized tensor (generated by Algorithm~\ref{sub_channel_clipping}) is shown in Figure~\ref{fig:sub_channel_per_tensor_clip}. As we can see, the combination of sub-channel quantization with greedy search over clipping parameter allows us to further reduce quantization error. In this example, the de-quantized tensor has only two numbers that are different from the input tensor. In Figure~\ref{fig:error_graph}, we show the quantization error of these approaches: 4 sub-channels with clipping (solid red line) and 8 sub-channels with clipping (solid green line). We observe that sub-channel quantization with clipping has the lowest quantization error, and we hypothesize that this approach will be useful for low bit quantization aware training.

\begin{algorithm}
  \caption{Sub channel quantization with clipping}\label{sub_channel_clipping}
  \begin{algorithmic}[1]
    \Procedure{QuantizeSubChClip}{$w$}\Comment{Input weights}
      \State $input\_shape\gets w.shape$ \Comment{Weights shape}
      \State $w\_sub\gets reshape(w)$ \Comment{Split into sub channels}
      \State $w\_q, scale, min\_val\gets greedy\_search(w\_sub)$
      \State $w\_deq \gets dequantize(w\_q, scale, min\_val)$
      \State $w\_deq \gets reshape(w\_deq, input\_shape)$
      \State \textbf{return} $w\_deq$\Comment{De-quantized weights}
    \EndProcedure
  \end{algorithmic}
\end{algorithm}

\begin{figure}[t]
  \centering
\begin{lstlisting}[language=python]
@tf.custom_gradient
def round(x):
  # Use STE for gradient.
  return tf.math.floor(x + 0.5), lambda dy: dy

def round_clip(x, prec):
  x = round(x)
  x = tf.clip_by_value(x, 0, 2.**prec - 1)  
  return x

def scale_and_min(x, prec, axis, clipping = 1.0, stop_gradient_scale=False):
  min_val = tf.math.reduce_min(x, axis=axis, keepdims=True)
  max_val = tf.math.reduce_max(x, axis=axis, keepdims=True)

  min_val = tf.multiply(min_val, clipping)
  max_val = tf.multiply(max_val, clipping)
  scale = tf.divide(max_val-min_val,2.**prec-1)
  
  if stop_gradient_scale: # STE over scale
    scale = tf.stop_gradient(scale)
  return scale, min_val

def quantize(x, prec, axis, clipping=1.0, stop_gradient_scale=False):
  scale, min_val = scale_and_min(x, prec, axis, clipping, stop_gradient_scale)
  x = x - min_val
  x = tf.math.divide_no_nan(x, scale)
  qx = round_clip(x, prec)
  return qx, scale, min_val

def dequantize(qx, scale, min_val):
  deqx = qx * scale
  deqx = deqx + min_val  
  return deqx
\end{lstlisting}  
  \caption{Quantization functions}
  \label{fig:quant_functions}
  \vspace{-4mm}
\end{figure}

\section{Experimental setups}

\subsection{Datasets}

Similar to~\cite{ASR4BITS}, we use LibriSpeech~\cite{Librispeech} to conduct QAT experiments. The LibriSpeech training set contains 960 hours of speech, where 460 hours of them are “clean” speech and the other 500 hours are “noisy” speech. We use “dev-clean" data to select the best model and then report its accuracy on “test-clean" and “test-other" data sets.



In terms of the experiments with large-scale datasets, we train the models with an in-house training set consisting of $\sim$1000 million United States English audio-text pairs from multiple domains, such as YouTube, search, and dictation. A  small  portion  of  the dataset is anonymized and hand-transcribed, while the rest is pseudo-transcribed with a 600M-parameter teacher system~\cite{Seong22}. Word  error  rates are  reported on 5.5K  anonymized  and hand-transcribed utterances representative of voice search traffic.

\subsection{Details in Conformer model architecture}

We use the same state-of-the-art Conformer Transducer~\cite{gulati2020conformer} backbones as in~\cite{ASR4BITS} for experiments on LibriSpeech and large-scale datasets.
For LibriSpeech experiments, the model has a single encoder with different number of layers for \textit{Small} (16 layers, 10M parameters) and \textit{Large} (17 layers, 118M) models. The decoder is a standard RNN-Transducer decoder with 1-layer LSTM.
The backbone of the experiments with large-scale data is based on~\cite{ding22b_interspeech}, consisting of a 7-layer causal conformer encoder (23-frame left context) and a 6-layer non-causal encoder (additional 30-frame right context). Each RNN-T decoder is comprised of an embedding prediction network and a fully-connected joint network.




\section{Results}


\subsection{Experiments on LibriSpeech}
\label{sec:exp_libri}

\begin{table}[t]
\begin{center}
\caption{Results of our proposed int2 QAT on Conformer Large(L) and Small(S) models with the baseline approaches on LibriSpeech test-clean and test-other subsets. Please see Section~\ref{sec:exp_libri} for the meanings of the method abbreviations.}
\label{table:librispeech}
\resizebox{\columnwidth}{!}{
  \begin{tabular}{l|cc|cc|cc}  
  \hline
  \multicolumn{4}{c}{Conformer (L)} \\
  \hline
  \textbf{Method} & \textbf{test-clean} & \textbf{test-other} & \textbf{Model size (MB)} \\
  \hline
  \cite{ASR4BITS} Float & \textbf{2.0} & \textbf{4.4} & 474.5 \\
  \hline
  I2Wsym & 3.6 & 8.1 & 53.8 \\
  I2Wasym & 2.2 & 5.0 & 54.0 \\
  I2WasymSc & 2.2 & 4.6 & 54.0 \\
  I2WasymScSubchClip & \textbf{2.0} & \textbf{4.5} & 55.3 \\ 
  \cite{ASR4BITS} I4W & \textbf{2.0} & \textbf{4.4} &  81.9 \\ 
  \cite{kim2021integer} I6W8A & 4.0 & 8.5 & 92.8 \\  
  \cite{kim2021integer} I8W & 3.1 & 7.1 & 123.7 \\  
  \cite{ASR4BITS} I8W & \textbf{2.0} & \textbf{4.5} & 138.1 \\

  \hline
  \hline
  \multicolumn{4}{c}{Conformer (S)} \\
  \hline
  Float & \textbf{2.5} & \textbf{6.1} & 41.5 \\
  \hline 
  I2Wsym & 8.7 & 19.6 & 10.1 \\
  I2Wasym & 4.3 & 10.3 & 10.2 \\
  I2WasymSc & 3.1 & 7.3 & 10.2 \\
  I2WasymScSubchClip & 3.1 & 7.0 & 10.5 \\
  \cite{ASR4BITS} I4W & 2.7 & 6.3 &  12.2 \\
  \cite{ASR4BITS} I8W & \textbf{2.5} & \textbf{6.0} &  16.4 \\
  
  \hline
  \hline
  \multicolumn{4}{c}{Other Architectures} \\
  \hline
  \cite{nguyen2020quantization} I8W  & 8.7 & 22.3 &  60 \\
  \cite{nguyen2020quantization} I6W  & 8.9 & 22.8 &  45 \\
  \cite{prasad2020quantization} I8WA & 6.9 & —    &   8 \\

  \hline
  \end{tabular}
}
\vspace{-10pt}
\end{center}
\end{table}

We experiment with the Conformer \textit{Large} and \textit{Small} models to examine the behaviors of 2-bit QAT with different model sizes. We hypothesize that the larger the model size, the easier to quantize its weights with no accuracy loss. The Conformer ASR model size is dominated by the Conformer blocks of the encoder, which account for 95\% and 82\% of the \textit{Large} and \textit{Small} models' disk utilization, respectively. For small models, decoder quantization disproportionately impacts model quality, while for large models the theoretical benefit of using decoder quantization becomes increasingly negligible. Hence, we decided to emulate \cite{ASR4BITS}'s approach and implement encoder-only quantization. As in~\cite{ASR4BITS}, we additionally opted to exclude quantization of the depthwise convolutional layers, as their contribution to model size is negligible. All other weights in the encoder's Conformer blocks are quantized with 2 bits, and their disk utilization is calculated assuming weight packing of 0.25 bytes/param.

We evaluate several 2-bit QAT configurations to see the impact of symmetric, asymmetric quantization, backpropagation on scale, and sub-channel quantization with clipping:
\begin{itemize}
    \item \textit{Float}: float32 weight, float32 activation (baseline)
    \item \textit{I2Wsym}: int2 weight with symmetrical quantization (we applied the approach in \cite{ASR4BITS}).
    \item \textit{I2Wasym}: int2 weight with per-channel asymmetrical quantization, described in section~\ref{sec:asym}, with full straight through estimator (stop\_gradient\_scale is set to True) as in \cite{ASR4BITS}.    
    \item \textit{I2WasymSc}: int2 weight with per-channel asymmetrical quantization, described in section~\ref{sec:asym_sc} and partial straight through estimator: with stop\_gradient\_scale is set to False, so that scale is part of the backpropagation.
    \item \textit{I2WasymScSubchClip}: int2 weight with per-channel asymmetrical quantization combined with sub-channel and clipping greedy search presented as Algorithm~\ref{sub_channel_clipping} and described in section~\ref{sec:asym_sc_sub_clip}. As \textit{I2WasymSc}, it uses partial straight through estimator: with stop\_gradient\_scale is set to False, so that scale is part of the backpropagation. We use 4 sub-channels for Conformer (L) and 8 for Conformer (S). Clipping greedy search was done in a  range from 0.8 to 1.0 with 0.02 step.
\end{itemize}

\noindent

We train Conformer(L) and Conformer(S) models with all above QAT configurations on 64 TPUs. Conformer(L) model converges with 150k training steps (it takes 2.3 days). Conformer(S) model converges with 400k steps (it takes 4.8 days). Note that QAT with \textit{I2WasymScSubchClip} is 40\% slower than other models due to the greedy search algorithm, but other presented approaches have no impact on training speed). In Table~\ref{table:librispeech}, we report the word error rate(WER) of Conformer(L) and Conformer(S) models trained with quantization techniques: \textit{I2Wsym}, \textit{I2Wasym}, \textit{I2WasymSc} and \textit{I2WasymScSubchClip}.
QAT with asymmetric quantization (\textit{I2Wasym}) allows us to use all four quantization buckets and reduce WER by 3\% absolute (in comparison to \textit{I2Wsym}) on Conformer(L) “test-other" data. QAT with asymmetric quantization and enabled backpropagation over scale (\textit{I2WasymSc}) further reduces WER by 0.4\% absolute on Conformer(L) “test-other" data. Combining \textit{I2WasymSc} with sub-channel and clipping greedy search \textit{I2WasymScSubchClip} gives the same WER with the state of the art int8 weights quantization model~\cite{ASR4BITS}, which is only 0.1\% worse than the float baseline on Conformer(L) "test-other" data. On 2-bit ASR model quantization we observe that the larger the model, the easier it is to quantize its weights with minimal accuracy loss.

\subsection{Exploring the limit of 2-bit quantization on large data}
\label{sec:exp_prod}

\begin{table}[t]
\begin{center}
\caption{Results of applying int2 QAT to production ASR model on large-scale data.}
\label{table:prod_model}
\resizebox{\columnwidth}{!}{
\begin{tabular}{c|l|c|c}
\hline
Exp & Model & WER & Model size (MB) \\
 \hline
 B0 & float32 model & 6.0 & 480 \\
 B1 & int4 & 6.3 & 65 \\
  \hline
 E0 & int2 & 12.6 & 37.5\\
 E1 & int2 I2WAsymScSubchClip & 7.6 & 37.5\\
 E2 & int2 I2WAsymScSubchClip + MSQE~\cite{quant_regularization} & 7.4 & 37.5\\

\hline
\end{tabular}}
\vspace{-20pt}
\end{center}
\end{table}

As shown in Table~\ref{table:prod_model}, when training the model with large-scale datasets, 4-bit quantization~\cite{ASR4BITS} (\textit{B1}: 6.3) can mostly retain the float model (\textit{B0}: 6.0) performance with minimal regression, which corresponds to the observations in~\cite{ASR4BITS}. However, quantizing the model to 2-bit becomes increasingly challenging, as the model is usually under-fitting. If we simply apply ASR quantization from ~\cite{ASR4BITS} for 2-bit QAT to the model (\textit{E0}), there is a significant WER regression compared to the float model (12.6 vs. 6.0). Alternatively, if we train the model with our best setup obtained from Section~\ref{sec:exp_libri} (\textit{E1}; i.e., asymmetric quantization, backpropagation on scale, and sub-channel quantization with clipping), the WER has been from 12.6 to 7.6 over the naive 2-bit QAT version \textit{E0}. In addition, we add an extra quantization regularization MSQE~\cite{quant_regularization} term \textit{E2} as MSE between weights and de-quantized weights. It can further mitigate the gap between 2-bit and float model by 0.2\%. In summary, our best performance 2-bit model \textit{E2} has a 1.4\% WER gap compared to the float model but with over 90\% of model size saving, compared to the float model. In terms of the 4-bit model \textit{B1}, the 2-bit model \textit{E2} has a 1.1\% WER gap but achieves over 40\% of model size reduction.

\section{Conclusion}

We proposed a novel approach of 2-bit QAT based on dynamic asymmetrical sub-channel quantization with adaptive per channel clipping. We reduced model size down to 55MB with minimal or no accuracy loss and established state of the art ASR model in terms of model size and WER.
We also showed that the larger the model ($>$ 100M parameters), the easier it is to quantize its weights with 2bit with no accuracy loss (it is important for large speech models).
When training the model with large-scale datasets, we illustrated the inevitable WER regression from the 2-bit model, and we showed that our proposed techniques can significantly mitigate the gap between 2-bit model and the float counterpart.

\bibliographystyle{IEEEtran}
{
\bibliography{mybib}

\begin{thebibliography}{10}
\providecommand{\url}[1]{#1}
\csname url@samestyle\endcsname
\providecommand{\newblock}{\relax}
\providecommand{\bibinfo}[2]{#2}
\providecommand{\BIBentrySTDinterwordspacing}{\spaceskip=0pt\relax}
\providecommand{\BIBentryALTinterwordstretchfactor}{4}
\providecommand{\BIBentryALTinterwordspacing}{\spaceskip=\fontdimen2\font plus
\BIBentryALTinterwordstretchfactor\fontdimen3\font minus
  \fontdimen4\font\relax}
\providecommand{\BIBforeignlanguage}[2]{{%
\expandafter\ifx\csname l@#1\endcsname\relax
\typeout{** WARNING: IEEEtran.bst: No hyphenation pattern has been}%
\typeout{** loaded for the language `#1'. Using the pattern for}%
\typeout{** the default language instead.}%
\else
\language=\csname l@#1\endcsname
\fi
#2}}
\providecommand{\BIBdecl}{\relax}
\BIBdecl

\bibitem{li2020comparison}
J.~Li \emph{et~al.}, ``On the comparison of popular end-to-end models for large
  scale speech recognition,'' in \emph{Interspeech 2020, 21st Annual Conference
  of the International Speech Communication Association}.\hskip 1em plus 0.5em
  minus 0.4em\relax {ISCA}, 2020, pp. 1--5.

\bibitem{CC18}
C.~Chiu \emph{et~al.}, ``State-of-the-art speech recognition with
  sequence-to-sequence models,'' in \emph{2018 {IEEE} International Conference
  on Acoustics, Speech and Signal Processing, {ICASSP}}, 2018.

\bibitem{KimHoriWatanabe17}
S.~Kim, T.~Hori, and S.~Watanabe, ``Joint ctc-attention based end-to-end speech
  recognition using multi-task learning,'' in \emph{2017 {IEEE} International
  Conference on Acoustics, Speech and Signal Processing, {ICASSP} 2017}.\hskip
  1em plus 0.5em minus 0.4em\relax {IEEE}, 2017, pp. 4835--4839.

\bibitem{JinyuLi2019}
J.~Li, R.~Zhao, H.~Hu, and Y.~Gong, ``Improving {RNN} transducer modeling for
  end-to-end speech recognition,'' in \emph{{IEEE} Automatic Speech Recognition
  and Understanding Workshop, {ASRU} 2019}.\hskip 1em plus 0.5em minus
  0.4em\relax {IEEE}, 2019, pp. 114--121.

\bibitem{gulati2020conformer}
A.~Gulati \emph{et~al.}, ``Conformer: Convolution-augmented transformer for
  speech recognition,'' in \emph{Interspeech 2020, 21st Annual Conference of
  the International Speech Communication Association}.

\bibitem{strom1997sparse}
N.~Str{\"{o}}m, ``Sparse connection and pruning in large dynamic artificial
  neural networks,'' in \emph{Fifth European Conference on Speech Communication
  and Technology, {EUROSPEECH} 1997}.\hskip 1em plus 0.5em minus 0.4em\relax
  {ISCA}, 1997.

\bibitem{takeda2017node}
R.~Takeda, K.~Nakadai, and K.~Komatani, ``Node pruning based on entropy of
  weights and node activity for small-footprint acoustic model based on deep
  neural networks,'' in \emph{Interspeech 2017, 18th Annual Conference of the
  International Speech Communication Association}.

\bibitem{li2018compression}
C.~Li \emph{et~al.}, ``Compression of acoustic model via knowledge distillation
  and pruning,'' in \emph{24th International Conference on Pattern Recognition,
  {ICPR} 2018}.

\bibitem{han2015deep}
S.~Han, H.~Mao, and W.~J. Dally, ``Deep compression: Compressing deep neural
  network with pruning, trained quantization and huffman coding,'' in \emph{4th
  International Conference on Learning Representations, {ICLR} 2016}.

\bibitem{alvarez2016efficient}
R.~Alvarez, R.~Prabhavalkar, and A.~Bakhtin, ``On the efficient representation
  and execution of deep acoustic models,'' in \emph{Interspeech 2016, 17th
  Annual Conference of the International Speech Communication Association},
  2016.

\bibitem{quantize_weight_act}
I.~Hubara \emph{et~al.}, ``Quantized neural networks: Training neural networks
  with low precision weights and activations,'' \emph{J. Mach. Learn. Res.},
  2017.

\bibitem{he2019streaming}
Y.~He \emph{et~al.}, ``Streaming end-to-end speech recognition for mobile
  devices,'' in \emph{ICASSP 2019-2019 IEEE International Conference on
  Acoustics, Speech and Signal Processing (ICASSP)}.\hskip 1em plus 0.5em minus
  0.4em\relax IEEE, 2019, pp. 6381--6385.

\bibitem{sainath2020streaming}
T.~N. Sainath \emph{et~al.}, ``A streaming on-device end-to-end model
  surpassing server-side conventional model quality and latency,'' in
  \emph{2020 IEEE International Conference on Acoustics, Speech and Signal
  Processing (ICASSP)}.

\bibitem{POSTQUANT}
\BIBentryALTinterwordspacing
 [Online]. Available:
  \url{https://www.tensorflow.org/lite/performance/post_training_quantization}
\BIBentrySTDinterwordspacing

\bibitem{pact}
J.~Choi \emph{et~al.}, ``Pact: Parameterized clipping activation for quantized
  neural networks,'' 2018.

\bibitem{clipping1}
H.~Wu \emph{et~al.}, ``Integer quantization for deep learning inference:
  Principles and empirical evaluation,'' 2020.

\bibitem{abdolrashidi2021pareto}
A.~Abdolrashidi \emph{et~al.}, ``Pareto-optimal quantized resnet is mostly
  4-bit,'' in \emph{{IEEE} Conference on Computer Vision and Pattern
  Recognition Workshops, {CVPR} Workshops 2021}.

\bibitem{ASR4BITS}
S.~Ding \emph{et~al.}, ``4-bit conformer with native quantization aware
  training for speech recognition,'' in \emph{Interspeech 2022, 23rd Annual
  Conference of the International Speech Communication Association}.\hskip 1em
  plus 0.5em minus 0.4em\relax {ISCA}, 2022.

\bibitem{flt_quant}
N.~Wang \emph{et~al.}, ``Training deep neural networks with 8-bit floating
  point numbers,'' in \emph{Advances in Neural Information Processing Systems
  31: Annual Conference on Neural Information Processing Systems 2018, NeurIPS
  2018}.

\bibitem{hotword4bit}
Y.~Mishchenko \emph{et~al.}, ``Low-bit quantization and quantization-aware
  training for small-footprint keyword spotting,'' in \emph{18th {IEEE}
  International Conference On Machine Learning And Applications, {ICMLA} 2019},
  2019.

\bibitem{fasoli20214}
\BIBentryALTinterwordspacing
A.~Fasoli \emph{et~al.}, ``4-bit quantization of lstm-based speech recognition
  models,'' in \emph{Interspeech 2021, 22nd Annual Conference of the
  International Speech Communication Association, Brno, Czechia, 30 August - 3
  September 2021}.\hskip 1em plus 0.5em minus 0.4em\relax {ISCA}, 2021, pp.
  2586--2590. [Online]. Available:
  \url{https://doi.org/10.21437/Interspeech.2021-1962}
\BIBentrySTDinterwordspacing

\bibitem{parrot4bit}
O.~Rybakov \emph{et~al.}, ``Streaming parrotron for on-device speech-to-speech
  conversion,'' 2022.

\bibitem{sub8bits}
K.~Zhen \emph{et~al.}, ``Sub-8-bit quantization for on-device speech
  recognition: {A} regularization-free approach,'' in \emph{{IEEE} Spoken
  Language Technology Workshop, {SLT} 2022}.

\bibitem{low_bits_img0}
M.~Courbariaux \emph{et~al.}, ``Binaryconnect: Training deep neural networks
  with binary weights during propagations,'' in \emph{Advances in Neural
  Information Processing Systems 28: Annual Conference, 2015}.

\bibitem{low_bits_img}
O.~Shayer, D.~Levi, and E.~Fetaya, ``Learning discrete weights using the local
  reparameterization trick,'' in \emph{6th International Conference on Learning
  Representations, {ICLR} 2018}.

\bibitem{xnor_net}
M.~Rastegari \emph{et~al.}, ``Xnor-net: Imagenet classification using binary
  convolutional neural networks,'' in \emph{Computer Vision - {ECCV} 2016 -
  14th European Conference}, ser. Lecture Notes in Computer Science, 2016.

\bibitem{asr1bit2bit4bit}
C.-F. Yeh, W.-N. Hsu, P.~Tomasello, and A.~Mohamed, ``Efficient speech
  representation learning with low-bit quantization,'' 2023.

\bibitem{praxis}
``praxis: https://github.com/google/praxis.''

\bibitem{white_paper_quantization}
M.~Nagel \emph{et~al.}, ``A white paper on neural network quantization,'' 2021.

\bibitem{RAND}
D.~Qiu, D.~Rim, S.~Ding, O.~Rybakov, and Y.~He, ``Rand: Robustness aware norm
  decay for quantized seq2seq models,'' 2023.

\bibitem{sub-channel}
A.~Gholami, S.~Kim, Z.~Dong, Z.~Yao, M.~W. Mahoney, and K.~Keutzer, ``A survey
  of quantization methods for efficient neural network inference,'' 2021.

\bibitem{group_quant}
H.~Yu \emph{et~al.}, ``Low-bit quantization needs good distribution,'' in
  \emph{2020 {IEEE/CVF} Conference on Computer Vision and Pattern Recognition,
  {CVPR} Workshops}.

\bibitem{ptq_group}
Z.~Yuan \emph{et~al.}, ``Ptq-sl: Exploring the sub-layerwise post-training
  quantization,'' 2021.

\bibitem{clipping2}
C.~Sakr \emph{et~al.}, ``Optimal clipping and magnitude-aware differentiation
  for improved quantization-aware training,'' in \emph{International Conference
  on Machine Learning, {ICML} 2022}, ser. Proceedings of Machine Learning
  Research, vol. 162.\hskip 1em plus 0.5em minus 0.4em\relax {PMLR}, 2022.

\bibitem{Librispeech}
V.~Panayotov \emph{et~al.}, ``Librispeech: An asr corpus based on public domain
  audio books,'' in \emph{2015 IEEE International Conference on Acoustics,
  Speech and Signal Processing (ICASSP)}, 2015.

\bibitem{Seong22}
D.~Hwang \emph{et~al.}, ``Pseudo label is better than human label,'' in
  \emph{Interspeech 2022, 23rd Annual Conference of the International Speech
  Communication Association}.\hskip 1em plus 0.5em minus 0.4em\relax {ISCA},
  2022.

\bibitem{ding22b_interspeech}
S.~Ding \emph{et~al.}, ``A unified cascaded encoder {ASR} model for dynamic
  model sizes,'' in \emph{Interspeech 2022, 23rd Annual Conference of the
  International Speech}.\hskip 1em plus 0.5em minus 0.4em\relax {ISCA}, 2022,
  pp. 1706--1710.

\bibitem{kim2021integer}
S.~Kim \emph{et~al.}, ``Integer-only zero-shot quantization for efficient
  speech recognition,'' in \emph{{IEEE} International Conference on Acoustics,
  Speech and Signal Processing, {ICASSP} 2022}.

\bibitem{nguyen2020quantization}
H.~D. Nguyen, A.~Alexandridis, and A.~Mouchtaris, ``Quantization aware training
  with absolute-cosine regularization for automatic speech recognition,'' in
  \emph{Interspeech 2020, 21st Annual Conference of the International Speech
  Communication Association}.

\bibitem{prasad2020quantization}
A.~Prasad, P.~Motlicek, and S.~Madikeri, ``Quantization of acoustic model
  parameters in automatic speech recognition framework,'' \emph{arXiv preprint
  arXiv:2006.09054}, 2020.

\bibitem{quant_regularization}
Y.~Choi, M.~El-Khamy, and J.~Lee, ``Learning sparse low-precision neural
  networks with learnable regularization,'' \emph{{IEEE} Access}, vol.~8, pp.
  96\,963--96\,974, 2020.

\end{thebibliography}
}

\end{document}